\documentclass[prl,twocolumn,amsmath,showpacs]{revtex4}
\usepackage{bm}
\usepackage{graphicx}

\begin{document}

\title{Theoretical Evidence for the Berry-Phase Mechanism of Anomalous
Hall Transport:\\ First-principles Studies on
CuCr$_2$Se$_{4-x}$Br$_x$}

\author{Yugui Yao$^{1}$, Yongcheng Liang$^{5,1}$, Di Xiao$^{2}$, Qian Niu$^{2}$,
 Shun-Qing Shen$^{3}$, X. Dai$^{1,3}$, and Zhong Fang$^{1,4}$}

\affiliation{$^1$ Beijing National Laboratory for Condensed Matter
Physics, Institute of Physics, Chinese Academy of Science, Beijing,
100080, China}

\affiliation{$^2$ Department of Physics, The University of Texas,
Austin, Texas, 78712-0264, USA}

\affiliation{$^3$ Department of Physics, and Center of Theoretical
and Computational Physics, The University of Hong Kong, Hong Kong,
China}

\affiliation{$^4$ International Center for Quantum Structure,
Chinese Academy of Sciences, Beijing, 100080, China}

\affiliation{$^5$ Insititute of Nano Science, Nanjing University of
Aeronautics and Aetronautics, Nanjing 210016, China}


\date{\today }

\begin{abstract}
To justify the origin of anomalous Hall effect (AHE), it is highly
desirable to have the system parameters tuned continuously. By
quantitative calculations, we show that the doping dependent sign
reversal in CuCr$_{2}$Se$_{4-x}$Br$_{x}$, observed but not understood,
is nothing but direct evidence for the Berry-Phase mechanism of
AHE. The systematic calculations well explain the experiment data for
the whole doping range where the impurity scattering rates is changed
by several orders with Br substitution. Further sign change is also
predicted, which may be tested by future experiments.
\end{abstract}

\maketitle

In spite of the wide applications of anomalous Hall effect (AHE) to
characterize ferromagnetism, its origin has been a controversial
subject since its discovery more than a century ago~\cite{hall1880}.
The {\bf k}-space gauge fields, known as the Berry curvature, exist
ubiquitously in Bloch bands where time reversal symmetry is broken,
giving rise to an intrinsic anomalous Hall effect (AHE) in
ferromagnets~\cite{NIU}.  This intrinsic effect was originally derived
by Karplus-Luttinger fifty years ago based on a linear response
theory~\cite{luttinger1954}, but was disputed ever since and until
recently, extrinsic mechanisms of skew scattering and side jump were
usually invoked~\cite{smit1955}.  Inspired by the new understanding
from the Berry phase connection~\cite{Nagaosa,jungwirth,haldane}, a
number of quantitative studies have been successfully carried out in
recent years~\cite{Fang2003,yao04}, finding that the Berry-phase
mechanism is important in various materials.  However, theoretical
understanding of the condition for such importance is far from clear,
despite a large number of theoretical analysis based on model
Hamiltonians~\cite{Nagaosa2}.  To fully explore the importance of the
Berry-phase mechanism, it is highly desirable to have a systematic
study of real materials in comparison to experiments when the system
parameters are tuned continuously.

In this paper, we report systematical first-principles calculations on
doping-dependence of the intrinsic AHE. Our material of choice is the
ferromagnetic spinel, CuCr$_2$Se$_4$, one of the parent compounds of a
wide class of colossal magnetoresistive chalcospinels.  It is well
known for its high Curie temperature ($T_c = 450$ K) and large
room-temperature magneto-optic Kerr effect, with great potential for
spintronics applications~\cite{Ramesha2004}. The experimental
measurement of AHE in this compound has been recently carried out by
Lee \textit{et al}.~\cite{Lee2004a}, where they are able to tune the
scattering rate by 70 folds with Br substitution of Se.  Our
quantitative calculations well explain the experimental AHE data over
the whole doping range with reasonable accuracy based on the
Berry-Phase mechanism. In particular, we reveal that the sharp sign
change in the doping dependent anomalous Hall conductivity, which was
observed in the experiment but not discussed explicitly, is a direct
evidence for the Berry-Phase mechanism of AHE. The sign change is due
to a large patch of high Berry-curvature in the band structure.  In
addition to explaining this experiment, our calculations also extend
to the case of hole doping, urging further experiments on the spinel
system.

The spin-polarized ground state of CuCr$_2$Se$_4$ has been calculated
by the X$\alpha $ method~\cite{Ogata1982} and by the linearized
muffin-tin orbital method~\cite{Ramesha2004,Antonov1999}.  In this
work, the relativistic electronic structure is calculated
self-consistently using the full-potential linearized augmented
plane-wave method with generalized gradient approximation
(GGA)~\cite{Perdew} for the exchange-correlation potential.  We use
the experimental lattice constant, and the muffin-tin radius
$R_{MT}=2.1,2.4,2.3$ Bohr for Cu, Cr, and Se atoms, respectively.  The
convergence of present calculations has been well checked.

\begin{figure}[tbp]
\includegraphics[width=7.5cm]{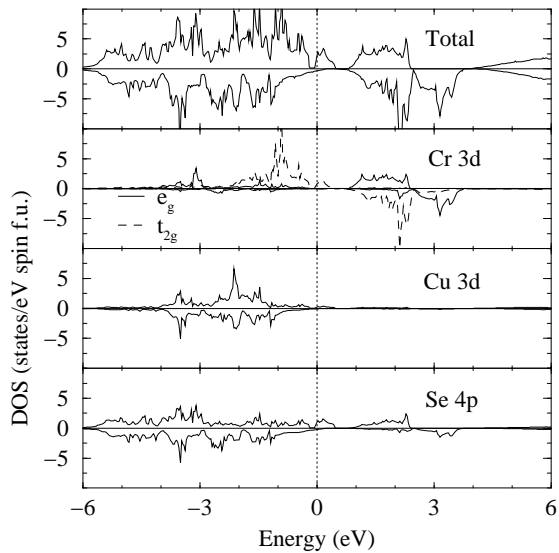}
\caption{\label{fig1} The calculated total and projected density of
states of CuCr$_2$Se$_{4}$. The Fermi level locates at energy zero.}
\end{figure}

Figure~\ref{fig1} shows the calculated total and projected density of
states of the parent compound CuCr$_2$Se$_4$, where Cr atoms occupy
the octahedral sites, and Cu atoms occupy the tetrahedral sites.  To
understand the complicated electronic structure, we consider the
compound as a combination of two parts: the tetrahedral (CuSe$_4$)
clusters (in the 6+ nominal valence) which are arranged periodically
in the crystal space of the diamond structure, and the Cr atoms (with
3+ nominal valence) in the interstitial sites of the CuSe$_4$
(diamond) crystal framework.  As shown in Fig.~\ref{fig1}, the
electronic states of (CuSe$_{4}$) is almost non-spin-polarized (the
slight polarization will be discussed later).  The Cu is nearly in the
Cu$^{+}$ valence state, whose $3d$ orbitals are almost fully occupied
and are away from the Fermi level.  The electronic states around the
Fermi level mostly come from the Cr $3d$ and Se $4p$ states.  The $3d$
states of Cr$^{3+}$ are exchange split by about 3.0 eV, giving rise to
the high spin configuration ($t_{2g}^{3\uparrow} e_g^{0\uparrow}
t_{2g}^{0\downarrow} e_g^{0\downarrow}$) with 3.0 $\mu_{B}$/Cr local
moment.  Here the Cr $3d$$-$Se $4p$ hybridization is an essential
factor to form the final electronic structure.  First, the
hybridization will induce holes in the Cr $t_{2g}^{3\uparrow}$ states,
resulting in reduced local moment and enhanced valence
(Cr$^{3+\delta}$).  This is evident from the slightly non-occupation
of Cr $t_{2g}^{\uparrow}$ states around the Fermi level (see
Fig.~\ref{fig1}).  Second, the hybridization leads to the negative
spin polarization of itinerant Se $4p$ states (anti-parallel to the
spin moment of Cr), which is crucial for the AHE in this compound.
Finally, the hybridization stabilizes the ferromagnetic ground state
and contributes to the high Curie temperature as discussed for
SrFeMoO$_6$ and (GaMn)As~\cite{SFMO}.  The calculated total moment is
5.1 $\mu _{B}$/f.u.\ for the parent compound, which is in good
agreement with the experimental value of 5.2
$\mu_B$/f.u~\cite{Lee2004a}.

The Br substitution will introduce additional electrons (in addition
to increasing disorder) due to the reduced negative valence of Br
compared with Se. It is justified by the following facts that the
electronic structure with doping can be described by the rigid-band
shift ({\it i.e.} changing doping is equivalent to sweeping the Fermi
energy) without losing the main physics for our purpose. (1) It was
reported~\cite{Lee2004a} that the Br substitution only affects the
Curie temperature, but does not affect the ferromagnetic ground state
dramatically.  (2) By 25\% substitution ($x=1.0$), the lattice
parameter changes only by 0.7\%~\cite{lattice}.  (3) As a
self-consistent check, the obtained electronic structures with the
rigid-band approximation is used to calculate the magnetic moments and
gives results in good agreement with experimental data.  As shown in
Table~\ref{tbl1}, the calculated total moment per f.u. increases
monotonically from 5.1~$\mu_B$ for $x=0.0$ to 6.0~$\mu_B$ for $x=1.0$,
while the experiment shows an increase from 5.2~$\mu_B$ to
6.0~$\mu_B$~\cite{Lee2004a}.  The calculated spin and orbital moments
of each atom also agree well with the results by X-ray magnetic
circular dichroism studies~\cite{Ramesha2004}.  The orbital moment of Se sites
mainly comes from its $4p$ states due to the spin-orbit coupling.

\begin{table}[tbp]
\caption{\label{tbl1}The calculated spin, orbital and total moments of
CuCr$_2$Se$_{4-x}$Br$_{x}$ in unit of $\mu_B$.}
\doublerulesep 0pt
\begin{tabular}{|c|c|c|c|c|c|c|c|}
\hline
\raisebox{-1.5ex}[0pt]{x} & \multicolumn{3}{|c|}{Orbital Moment/site} &
\multicolumn{3}{|c|}{Spin Moment/site} & \raisebox{-1.5ex}[0pt]{Total/f.u.}
\\ \cline{2-4}\cline{5-7}
& Cu & Cr & Se & Cu & Cr & Se &  \\
 \hline 0.0 & -0.010 & -0.0096 & -0.0030 & -0.12 & 2.80 & -0.16 & 5.08 \\
 \hline 0.2 & -0.0060 &-0.0077 & -0.0046 & -0.11 & 2.84 & -0.15 & 5.23 \\
 \hline 0.4 &-0.0017 & -0.0045 & -0.0067 & -0.091 & 2.87 & -0.13 & 5.38 \\
 \hline 0.6 & -0.0002 & -0.0027 & -0.0066 & -0.072 & 2.91 & -0.11 & 5.61 \\
  \hline 0.8 & 0.0012 & 0.0012 & -0.0063 & -0.049 & 2.93 & -0.095 & 5.80 \\
  \hline 1.0 & -0.0019 & 0.0046 & -0.0033 & -0.022 & 2.95 & -0.072 & 5.99 \\
   \hline
\end{tabular}%
\end{table}

The intrinsic anomalous Hall conductivity can
be evaluated from the linear response theory using the standard Kubo
formula~\cite{yao04}
\begin{equation}
\sigma_{xy}
=\frac{e^{2}}{\hbar}\int\frac{d^3k}{(2\pi)^3}\sum_{n}f_{nk}
\bm \Omega_{n}^{z}(\bm k)
\end{equation}
with
\begin{equation}
\bm\Omega_{n}^{z}(\bm k)=\sum_{n'\neq n}\frac{2\mathrm{Im}
\langle \psi_{nk}|v_x|\psi_{n'k}\rangle
\langle \psi_{n'k}|v_y|\psi_{nk}\rangle}
{(\omega_{n'k}-\omega_{nk})^2-(i\delta)^2}
\end{equation}
where $|\psi_{nk}\rangle$ is the eigenstate with eigenvalue
$E_{nk}=\hbar \omega _{nk}$, $v_{x}$ and $v_{y}$ are the velocity
operators, $f_{nk}$ is the Fermi-Dirac distribution function, and
$\delta$ is a small parameter representing the finite life-time
broadening of the eigenstates.  The $\bm\Omega_n(\bm k)$ is a
vector in the $\bm k$-space, and can be related to the Berry
curvature of the Bloch state in the clean limit ($\delta=0$),
i.e., $\bm\Omega_n(\bm k) = \mathrm{Im} \langle \nabla_{\bm
k}u_{n\bm k}|\times|\nabla_{\bm k}u_{n\bm k}\rangle$ with $u_{n\bm
k}$ being the periodic part of the Bloch wave function.

\begin{figure}[tbp]
\includegraphics[width=7.5cm]{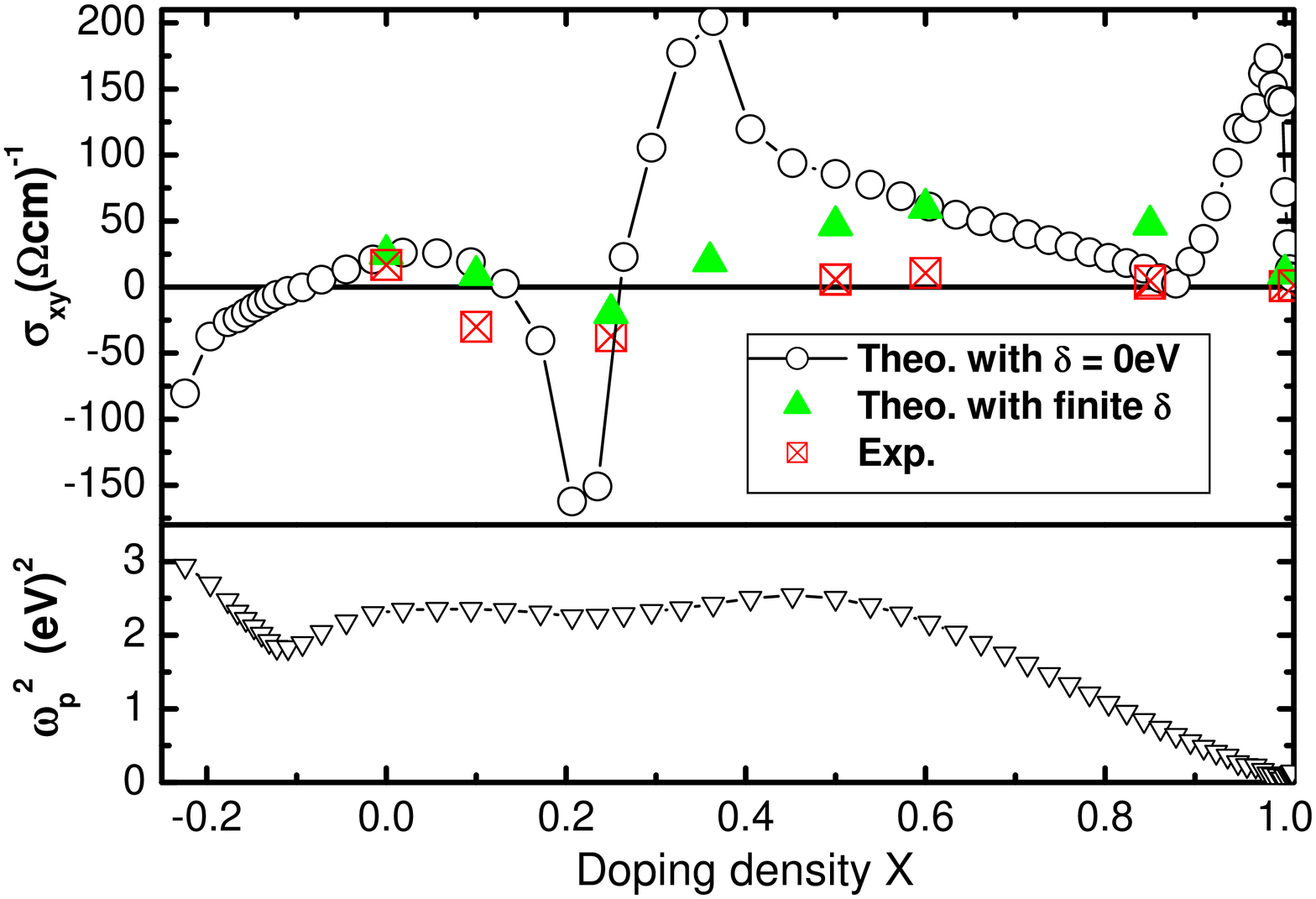}
\caption{\label{fig2}(Color online). The anomalous Hall conductivity
$\sigma_{xy}$ as function of doping $x$ in
CuCr$_2$Se$_{4-x}$Br$_x$. For the theoretical results, the open
circles are $\protect\sigma_{xy}$ for the clean limit ($\delta$=0),
and the triangles are $\sigma_{xy}$ with finite $\delta$ in the Kubo
formula, where the doping dependent $\delta$ is determined from the
\textit{ab-initio} calculated plasma frequency (shown in the lower
panel) and the experimental longitudinal resistivity (see the text
part for details). The square-crosses are experimental results from
Ref.~\cite{Lee2004a}.}
\end{figure}

Figure~\ref{fig2} shows the calculated intrinsic $\sigma_{xy}$ as a
function of the doping $x$.  Let us first consider the calculated
curve in the clean limit ($\delta=0$, open circles). It is obvious
that $\sigma_{xy}$ is highly non-monotonic and changes its sign twice
between $x=0.0$ and $0.5$; it starts with a positive value at $x=0$,
then changes its sign to negative around $x=0.1$, and again to
positive for $x>0.3$.  The places where $\sigma_{xy}$ changes sign
with varying $x$, although appear arbitrary, are in good agreement
with the experimental data (the square-cross points in
Fig.~\ref{fig2}).  While a quantitative comparison with the experiment
of the overall behavior of $\sigma_{xy}$ needs further analysis (as
addressed below), such an agreement is a striking result, considering
the fact that the calculations were done systematically for the whole
doping region without adjustable parameters.  In the experimental
analysis~\cite{Lee2004a}, the sign of $\sigma_{xy}$ is dropped; only
its absolute value is taken into account.  Our result here, however,
shows that the sign of $\sigma_{xy}$ is important, and the sign change
of $\sigma_{xy}$ with varying doping $x$ is a natural result of the
Berry-phase mechanism of AHE.

To make the quantitative comparison, we need consider the effect of
the finite life-time of the eigenstates.  The simplest way to do this
is to assume the diagonal form of electron self-energy and to use a
single parameter $\delta$ instead (in the Kubo formula), thereby
neglecting the vertex correction due to impurity scattering. It is
worth to note that the doping dependent $\delta$ in our approach is
not adjustable parameter but instead it is determined from the
relaxation time $\tau=\hbar/\delta=1/(\varepsilon_0\omega_p^2\rho)$,
where $\rho$ is the longitudinal resistivity, adopted from
Ref.~\cite{Lee2004a}, and the plasma frequency $\omega_p$ is
calculated from the band structure by
\[
\omega_{p}^{2}=\frac{e^2}{\pi ^2 m^2}\sum_n\int d^3k \langle
\psi_{nk}|p_x|\psi_{nk}\rangle \langle \psi_{nk}|p_x |\psi
_{nk}\rangle \delta (\varepsilon _{n\bm k}-\varepsilon_F).
\]
The plasma frequency is actually the measurement of the ratio
between the number of band carriers $n^*$ and the effective mass
of electrons $m^*$, according to the relation $\omega_p^2=
n^*e^2/(\varepsilon_0 m^*$).  The triangle points in
Fig.~\ref{fig2} are the theoretical values of $\sigma_{xy}$ after
considering the effect of relaxation.  It is now obvious that the
calculated intrinsic $\sigma_{xy}$ is in quantitative agreement
with experimental data, especially in the region around the sign
change ($x=0.3$).

\begin{figure}[tbp]
\includegraphics[width=7.5cm]{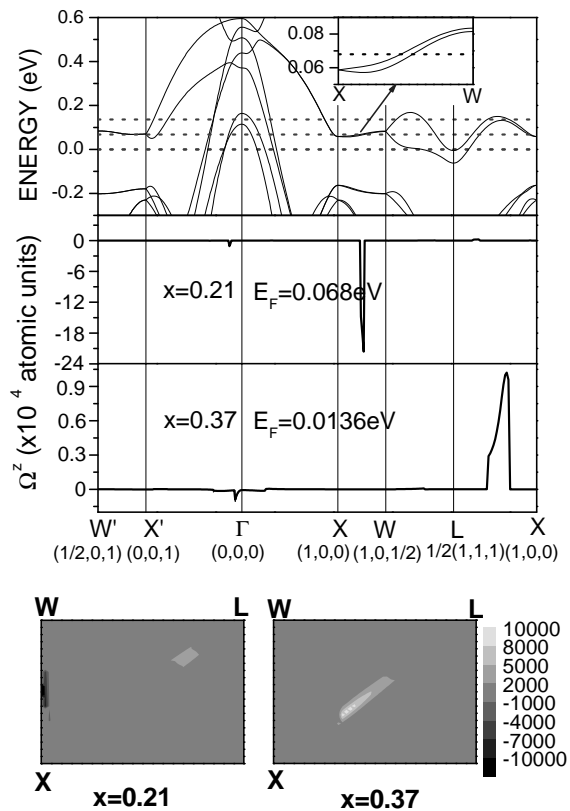}
\caption{\label{fig3} The calculated band structure of
CuCr$_2$Se$_{4-x}$Br$_x$ (upper panel) and the sum of Berry
curvature over the occupied bands $\Omega^z(\bm k)=\sum_n
f_{nk}\Omega_n^z(\bm k)$ for two characteristic Fermi level
positions corresponding to doping $x=0.21$ and $0.37$ respectively.
The lower panels show the $\Omega^z(\bm k) $ in a special plane of
the BZ for different doping $x$.}
\end{figure}

The doping-dependent sign change of $\sigma_{xy}$ was observed before
in other ferromagnetic alloys, such as Ni$_{x-x}$Fe$_x$, Au-Fe and
Au-Ni alloys~\cite{hall1880}.  The phenomenological
theory~\cite{Nozieres} relates the sign change to the change of
effective spin-orbit coupling with varying chemical potential.  Here
our numerical calculation indicates that the sign change in
CuCr$_2$Se$_{4-x}$Br$_x$ is attributed microscopically to the
topological nature of electronic bands in the Berry phase picture.
From simplified two-band mode, it is understood that the sum of Berry
curvatures over the occupied bands $\Omega^{z}(\bm k)=\sum_n
f_{nk}\Omega_n^z(\bm k)$ is spiky and the sign change occurs near the
degenerate or band crossing points, which act as magnetic monopoles in
the momentum space~\cite{Fang2003}. As a result, by summing over the
Brillouin zone (BZ), $\sigma_{xy}$ is typically a non-monotonic
function of chemical potential, and exhibits sharp fluctuation.  This
is the case for CuCr$_2$Se$_{4-x}$Br$_x$ as shown in
Fig.~\ref{fig3}. A similar behavior was also observed in the
2-dimensional (2D) systems, such as the sign change in the quantum
well structure~\cite{Dai}. However, we note that the higher
dimensionality in the present system makes the problem quite
different.  In the 3D case, the single band crossing point cannot
contribute enough weight to the sign change of $\sigma_{xy}$ due to
the 3D (instead of 2D) integration of BZ.  To get enough weight, a
high density of states near band crossing points (or near degenerate
points) are necessary (for example, the insert in Fig.~\ref{fig3}).
Due to the presence of band dispersion, it is generally hard to have
all those band crossing points occupied (or unoccupied) at each fixed
chemical potential, which leads to lower possibility for the sign
change of $\sigma _{xy}$ in 3D than in the 2D case.  On the other
hand, CuCr$_2$Se$_{4-x}$Br$_x$ is an isotropic 3D system where sharp
sign changes of $\sigma_{xy}$ are observed. Actually, the sign changes
in CuCr$_2$Se$_{4-x}$Br$_x$ are neither from simple band crossing nor
from the high symmetric points of the BZ.  As shown in
Fig.~\ref{fig3}, the dominant negative Berry curvature for $x=0.21$
(the valley of $\sigma_{xy}$) and the positive Berry curvature for
$x=0.37$ (the peak of $\sigma_{xy}$) are located at different regions
of the BZ.  We have tried to use an effective Luttinger Hamiltonian
(fitted from our electronic structure calculations) to study the
system, but the sign changes cannot be reproduced even qualitatively.
This indicates that in realistic materials the accurate
first-principles calculations are important.

In conclusion, the doping-dependent AHE in CuCr$_2$Se$_{4-x}$Br$_x$
is investigated by \textit{ab initio} calculations, and analyzed
according to the Berry phase picture.  The good agreement between
experimental and numerical results provide strong evidence for the
Berry-phase mechanism of AHE, even when the impurity scattering
rates is changed by several orders. The disorder (extrinsic)
contributions, which may also be related to the non-zero Berry
curvature~\cite{disorder}, can change the magnitude of our
calculated AHE quantitatively, but they are not expected to affect
such features as the sign change qualitatively.  To further verify
our results, we point out the following two aspects which can be
justified experimentally.  (1) Additional sign change is predicted
from our calculation.  As shown in Fig.~\ref{fig2}, by negative
doping (hole doping), we predict that $\sigma_{xy}$ changes its sign
from positive to negative.  The hole doping can be realized
experimentally by doping As instead of Br.  (2) The experimentally
observed Nernst effect~\cite{Nernst} in the same compound
CuCr$_2$Se$_{4-x}$Br$_x$ can be also checked from the present
picture~\cite{xiao2006}.

We acknowledge valuable discussions with Junren Shi, and are
grateful to Wei-Li Lee for sharing the original experimental data
and for discussion.  This work was supported by the Knowledge
Innovation Project of the Chinese Academy of Sciences, the NSFC
under the grant No.~10404035, 10534030, 10674163 (Y.G.Y),
90303022, 60576058, 10334090 and 10425418 (Z.F.), by the NSF under
grant No.~DMR-0404252 and DMR-0606485 (D.X.), by the DOE under
grant No.~DE-FG03-02ER45958 (Q.N.), and RGC of Hong Kong under
Grant No.~HKU 7042/06P (S.Q.S.).

\end{document}